\documentclass[fleqn,usenatbib]{mnras}

\usepackage{newtxtext,newtxmath}

\usepackage[T1]{fontenc}

\DeclareRobustCommand{\VAN}[3]{#2}
\let\VANthebibliography\thebibliography
\def\thebibliography{\DeclareRobustCommand{\VAN}[3]{##3}\VANthebibliography}

\usepackage{graphicx}	
\usepackage{amsmath}	
\usepackage{textcomp}	
\usepackage[dvipsnames]{xcolor}
\usepackage{tikz}
\usetikzlibrary{arrows.meta,
                 shapes,
                chains,
                positioning,
                shapes.geometric}
\tikzstyle{arrow}=[draw, -latex]

\newcommand{\appropto}{\mathrel{\vcenter{
  \offinterlineskip\halign{\hfil$##$\cr
    \propto\cr\noalign{\kern2pt}\sim\cr\noalign{\kern-2pt}}}}}

\newcommand{\bfR}{{\bf R}}

\newcommand{\aid}{\textbf{\textit{a}}_\mathrm{id}}
\newcommand{\aidt}{\tilde{\textbf{\textit{a}}}_\mathrm{id}}
\newcommand{\atwob}{\textbf{\textit{a}}_\mathrm{2b}}
\newcommand{\ad}{\textbf{\textit{a}}_\mathrm{d}}
\newcommand{\aone}{\textbf{\textit{a}}_\mathrm{N}}

\newcommand{\de}{\mathrm{d}}

\newcommand\ba{\begin{eqnarray}}
\newcommand\ea{\end{eqnarray}}

\defcitealias{Goodman2001}{GR01}
\defcitealias{Rafikov2002}{Rafikov 2002}
\defcitealias{Cimerman2021}{Paper I}
\defcitealias{GT80}{GT80}

\title[Conservation laws in non-inertial frames]{Conservation laws in non-inertial frames
and non-conservation of energy of relative motion in two-body problem}

\author[R. R. Rafikov]{Roman R. Rafikov$^{1,2}$\thanks{E-mail: rrr@damtp.cam.ac.uk}\\
$^{1}$Department of Applied Mathematics and Theoretical Physics, University of Cambridge, Wilberforce Road, Cambridge CB3 0WA, UK\\
$^{2}$Institute for Advanced Study, Einstein Drive, Princeton, NJ 08540, USA}

\date{Accepted XXX. Received YYY; in original form ZZZ}

\pubyear{2024}

\begin{document}

\def\etal{et al.\ \rm}
\def\etal{et al.\ \rm}
\def\Fdw{F_{\rm dw}}
\def\Fdis{F_{\rm dw,dis}}
\def\Fnu{F_\nu}
\def\WD{\rm WD}

\label{firstpage}
\pagerange{\pageref{firstpage}--\pageref{lastpage}}
\maketitle

\begin{abstract}
The dynamics of systems of multiple gravitationally interacting bodies is often studied in a frame attached to one of the objects (e.g. a central star in a planetary system). As this frame is generally non-inertial, indirect forces appear in the equations describing the motion of bodies relative to the reference object. According to the convention adopted in celestial mechanics, the associated indirect acceleration is defined to be different for every object under consideration, whereas the gravitational coupling between each body and the reference object is described via the effective two-body potential, which does not obey the equivalence principle. Here we point out that a slightly different and more physically motivated definition of the indirect acceleration provides significant benefits when interpreting relative motion in a non-inertial frame. First, the indirect acceleration ends up being the same for all objects in the system. Second, the non-conservation of momentum, angular momentum, and energy of the whole system in a non-inertial frame naturally follow from the action of the indirect acceleration on the system as an external force. We also argue that the vis viva integral of the classical two-body problem should not be interpreted as a statement of energy conservation in a non-inertial frame attached to one of the bodies. The energy of relative motion is not conserved in this frame due to the work done on the two-body system by the indirect force. These results can be useful for interpreting dynamics in various astrophysical contexts, in particular the physics of disc-planet coupling.  
\end{abstract}





\section{Introduction}
\label{sec:intro}


When studying systems of multiple gravitationally-interacting bodies, it is often advantageous to consider their dynamics in the reference frame attached to one of the bodies and moving with it. This is routinely done in a variety of astrophysical situations such as the dynamics of planetary systems, of accretion disks in binaries, of protoplanetary discs perturbed by planets, and so on. Working in a co-moving frame gives rise to the so-called {\it indirect forces} in the description of the motion of bodies relative to the reference object. These forces account for the motion of the reference object caused by its gravitational attraction to other bodies in the system. 

While mathematical implications of the presence of indirect forces are well understood in celestial mechanics, physical interpretation of these forces in their conventional form can be tricky. In this pedagogical note, we propose a reformulation of the indirect forces (Section \ref{sec:alt-motion}), which is slightly different from the classical approach (recited in Section \ref{sec:indirect}) used to describe dynamics in a non-inertial frame centered on one of the bodies. We show that this reformulation leads, in particular, to a very straightforward interpretation of conservation laws in a non-inertial system (Section \ref{sec:cons}). This interpretation is one of the key motivations for the present work and will be used in a companion study \citep{Rafikov2025} of disc-planet interaction in a non-inertial frame. We also re-examine the classical two-body problem (Section \ref{sec:2body}) and demonstrate that some of its well-established aspects, e.g. the vis viva integral and its relation to energy conservation, take on a new meaning in light of the proposed reformulation. Mathematical details of our derivations can be found in the Appendix.


\section{Dynamics in a non-inertial frame and indirect forces: conventional view}
\label{sec:indirect}


Consider a system of $N$ masses $m_1,\dots,m_N$ interacting via Newtonian gravity with their locations in an inertial frame specified by vectors ${\bf r}_n$, $n=1,\dots,N$. The equation of motion for a mass $m_n$ under the gravitational influence of all other masses following from the Newton's second law reads 
\ba
\ddot{\bf r}_n=-G\sum\limits_{\substack{k=1 \\ k\neq n}}^N m_k\frac{{\bf r}_n-{\bf r}_k}{\vert{\bf r}_n-{\bf r}_k\vert^3}.
\label{eq:EoMgen}
\ea

Let us now choose one of the masses, say $m_1$ (without loss of generality as we can always re-label them), and consider the motion of all other masses in the non-inertial reference frame co-moving with $m_1$. To do this, we introduce a position vector $\bfR_n$ of the mass $m_n$ relative to the reference mass,
\ba
{\bf R}_n={\bf r}_n-{\bf r}_1.
\label{eq:R}
\ea  
Then, using equation (\ref{eq:EoMgen}), we can write down the equation of motion for $m_n$ in the frame co-moving with $m_1$ as \citep[e.g.][]{Tremaine2023}
\ba
\ddot\bfR_n=-G(m_1+m_n)\frac{{\bf R}_n}{R_n^3}-G\sum\limits_{\substack{k=2 \\ k\neq n}}^N m_k\left(\frac{{\bf R}_n-{\bf R}_k}{\vert{\bf R}_n-{\bf R}_k\vert^3}+\frac{{\bf R}_k}{R_k^3}\right),
\label{eq:EoMRgen}
\ea
with $R_n=|\bfR_n|$. This is a standard form in which these equations are cast in celestial mechanics, with individual terms grouped as
\ba
\ddot{\bf R}_n  = \atwob + \ad + \aidt\, ,
\label{eq:acc-split}
\ea  
and representing the {\it effective two-body} acceleration between $m_1$ and $m_n$,
\ba
\atwob =-G(m_1+m_n)\frac{{\bf R}_n}{R_n^3}=-\nabla_{\bfR_n}\Phi_\mathrm{2b}({\bf R}_n),
\label{eq:a2b}
\ea  
the {\it direct} acceleration of $m_n$ due to all other masses in the system (except $m_1$), 
\ba
\ad =-G\sum\limits_{\substack{k=2 \\ k\neq n}}^N m_k\frac{{\bf R}_n-{\bf R}_k}{\vert{\bf R}_n-{\bf R}_k\vert^3}
=-\nabla_{\bfR_n}\Phi_\mathrm{d}({\bf R}_n),
\label{eq:ad}
\ea 
and the {\it indirect} acceleration
\ba
\aidt =-G\sum\limits_{\substack{k=2 \\ k\neq n}}^N m_k\frac{{\bf R}_k}{R_k^3}
=-\nabla_{\bfR_n}\tilde\Phi_\mathrm{id}({\bf R}_n).
\label{eq:aidtilde}
\ea  
These accelerations are gradients with respect to ${\bf R}_n$ of the three distinct potentials --- effective two-body, direct, and indirect potentials --- given by
\ba
\Phi_\mathrm{2b}({\bf R}_n) &=& -\frac{G(m_1+m_n)}{R_n}.
\label{eq:Phi2b}\\
\Phi_\mathrm{d}({\bf R}_n) &=& -\sum\limits_{\substack{k=2 \\ k\neq n}}^N \frac{Gm_k}{\vert{\bf R}_n-{\bf R}_k\vert},
\label{eq:direct-pot}\\
\tilde\Phi_\mathrm{id}({\bf R}_n) &=& \sum\limits_{\substack{k=2 \\ k\neq n}}^N Gm_k\frac{{\bf R}_k\cdot {\bf R}_n}{R_k^3}.
\label{eq:indirect-pot}
\ea
When studying the motion of mass $m_n$, the indirect acceleration $\aidt$ defined in this way accounts for the non-inertial motion of the mass $m_1$ caused by its gravitational attraction to all other objects in the system {\it except $m_n$} itself. This definition makes both the indirect acceleration and indirect potential functions of $n$, i.e. they are both different for different objects in the system.

The decomposition given by equations (\ref{eq:acc-split})-(\ref{eq:indirect-pot}) of $\ddot{\bf R}_n$ into different contributions follows a standard convention adopted in celestial mechanics \citep[e.g.][]{Smart1953,Brou1961,Murray1999,Tremaine2023}. However, it is not ideal for a number of reasons. First, the indirect acceleration represented by the last sum in equation (\ref{eq:EoMRgen}) excludes the gravitational effect of mass $m_n$ on $m_1$, and there is no a priori physical reason why this should be the case. Second, let us for a moment consider the two-body problem, when $N=2$. Then one finds $\Phi_\mathrm{d}=\tilde\Phi_\mathrm{id}=0$ so that the equation (\ref{eq:EoMRgen}) for the motion of mass $m_2$ reduces to a classical result 
\ba
\ddot{\bf R}_2=-G(m_1+m_2)\frac{{\bf R}_2}{R_2^3}= -\nabla_{\bfR_2}\Phi_\mathrm{2b}({\bf R}_2),
\label{eq:EoM2body}
\ea
where $\bfR_2= {\bf r}_2-{\bf r}_1$. In the classical interpretation, the effective two-body potential $\Phi_\mathrm{2b}=-G(m_1+m_2)/R_2$ fully represents the gravitational coupling responsible for the evolution of $\bfR_2$, with no other forces (except gravity) present in the two-body system. However, $\Phi_\mathrm{2b}$ is not a true gravitational potential as it does not satisfy the equivalence principle: according to equation (\ref{eq:EoM2body}), the acceleration of a mass $m_2$ due to $m_1$ depends on the mass $m_2$ itself, which violates the equivalence principle. Same issue arises in the general $N$-body case, see equation (\ref{eq:Phi2b}). We will discuss two-body problem in greater depth in Section \ref{sec:2body}. 

These issues motivate us to explore other interpretations for the meaning of different terms in equation (\ref{eq:EoMRgen}), which we do next.


\section{An alternative interpretation of the dynamics in a non-inertial frame}
\label{sec:alt-motion}


One can formulate an alternative interpretation of motion in a non-inertial frame attached to $m_1$ using the original equation (\ref{eq:EoMRgen}). We simply re-write the latter as
\ba
\ddot{\bf R}_n=-Gm_1\frac{{\bf R}_n}{R_n^3}-
G\sum\limits_{\substack{k=2 \\ k\neq n}}^N m_k\frac{{\bf R}_n-{\bf R}_k}{\vert{\bf R}_n-{\bf R}_k\vert^3}-
G\sum\limits_{k=2}^N m_k\frac{{\bf R}_k}{R_k^3},
\label{eq:EoMRgen-alt}
\ea
re-assigning $-Gm_n {\bf R}_n/R_n^3$ from the first term in equation (\ref{eq:EoMRgen}) to the last sum there. Thus, the last term in equation (\ref{eq:EoMRgen-alt}) now includes the $k=n$ contribution, unlike the last sum in the equation (\ref{eq:EoMRgen}). One can see that it is equal to $-\ddot{\bf r}_1$ --- the opposite of the acceleration of mass $m_1$ caused by its gravitational attraction to {\it all} other ($n=2,\dots,N$) objects, {\it including} $m_n$. Naturally, the first two terms in (\ref{eq:EoMRgen-alt}) represent $\ddot{\bf r}_n$, in accordance with definition (\ref{eq:R}). 

If we define the classical Newtonian gravitational acceleration 
\ba
\aone =-Gm_1\frac{{\bf R}_n}{R_n^3}
\label{eq:a_1}
\ea  
and the {\it modified indirect} acceleration (cf. equation (\ref{eq:aidtilde}))
\ba
\aid=-\ddot{\bf r}_1=-G\sum\limits_{k=2}^N m_k\frac{{\bf R}_k}{R_k^3},
\label{eq:acc}
\ea  
then equation (\ref{eq:EoMRgen-alt}) becomes (instead of equation (\ref{eq:acc-split}))
\ba
\ddot{\bf R}_n=\aone + \ad + \aid,
\label{eq:EoMRgen-alt1}
\ea
where $\ad$ is still given by equation (\ref{eq:ad}). 

This equation enables a very straightforward interpretation of the force balance in a non-inertial frame of $m_1$: accelerated motion of $m_1$ causes the emergence of an acceleration $\aid$ in the frame co-moving with $m_1$, and this (reflex) acceleration acts {\it uniformly} on all other $n=2,\dots,N$ objects in the system \citep[see also][]{Crida2025}. As a result, the acceleration $\ddot{\bf R}_n$ experienced by $m_n$ is contributed to by the classical gravitational attraction of $m_n$ to all other objects in the system (second term $\ad$), including $m_1$ (first term $\aone$), as well the reflex acceleration $\aid$ (the same for all $n=2,\dots,N$ objects) due to the non-inertial motion of $m_1$. This simple physical interpretation is lacking in the conventional approach described in Section \ref{sec:indirect}.

As before, $\ad$ is still related to $\Phi_\mathrm{d}$ via equation (\ref{eq:ad}), whereas $\aone = -\nabla_{\bfR_n}\Phi_\mathrm{N}(\bfR_n)$, where
\ba
\Phi_\mathrm{N}({\bf R}_n) = -\frac{Gm_1}{R_n}
\label{eq:Phi1-usual}
\ea
is the usual Newtonian gravitational potential due to mass $m_1$, which is different from the effective two-body potential $\Phi_\mathrm{2b}$ defined by equation (\ref{eq:Phi2b}). Importantly, $\Phi_\mathrm{N}$ is the true gravitational potential satisfying the equivalence principle, unlike $\Phi_\mathrm{2b}$. 

Also, $\aid = -\nabla_{\bfR_n}\Phi_\mathrm{id}(\bfR_n)$, where the {\it modified indirect} potential
\ba
~~~\Phi_\mathrm{id}({\bf R}_n) =-\frac{Gm_n}{R_n}+ \sum\limits_{\substack{k=2 \\ k\neq n}}^N Gm_k\frac{{\bf R}_k\cdot {\bf R}_n}{R_k^3}
\label{eq:EoMRgen-pot}
\ea
accounts for the motion of mass $m_1$ caused by {\it all} other objects, {\it including $m_n$}. The first term in the right-hand side of (\ref{eq:EoMRgen-pot}) is different from the rest of the terms in the sum because the $k=n$ contribution to $\aid$ integrates over $\bfR_n$ differently from all other contributions. Because of that, the expression for $\Phi_\mathrm{id}$ is still unique for each object in the system, even though $\aid$ is the same for all of them. Inclusion or exclusion of the effect of $m_n$ on the motion of $m_1$ is the key difference between $\aidt$ and $\aid$ (or $\tilde\Phi_\mathrm{id}$ and $\Phi_\mathrm{id}$): the former does not account for it, whereas the latter does.  

Note that we can absorb $\aone$ into $\ad$ in equation (\ref{eq:EoMRgen-alt1}) by simply extending the summation in the definition (\ref{eq:ad}) to run from $k=1$ (rather than $k=2$) and remembering that $\bfR_1={\bf 0}$. However, in many problems of interest, the mass $m_1$ is dominant compared to other masses (i.e.  a planetary system around a star, or a system of a supermassive black hole with a nuclear star cluster around it), in which case it is natural to show its effect explicitly, while the gravity of all other objects can be considered as a perturbation.  

In such cases, when studying the dynamics of mass $m_n$, in a zeroth-order approximation one would set all $m_k=0$, $k= 2,\dots,N$, in the equation (\ref{eq:EoMRgen-alt}). In this limit $\ad=\aid={\bf 0}$, so that equation (\ref{eq:EoMRgen-alt1}) reduces to equation $\ddot{\bf R}_k=-(Gm_1/R_k^3){\bf R}_k$ describing relative motion of a test mass in the classical two-body problem. The neglected terms proportional to $m_k$, $k= 2,\dots,N$, would get added at the next level of approximation as a perturbation, as is normally done in celestial mechanics. 

Note that this procedure does not provide an exact solution of the two-body problem for each individual object $m_k$ ($k>1$) at the zeroth-order approximation. This is because the angular frequency of the test mass problem $n^2=Gm_1/a_k^3$ ($a_k$ is the semi-major axis of $m_k$) is slightly different from the two-body problem with a finite $m_k$, $n^2=G(m_1+m_k)/a_k^3$. To remedy this problem, one can employ a staged approach to adding masses $m_i$ into the consideration: for every mass $m_k$ one would first add only the terms involving $m_k$, keeping all other $m_i=0$ ($i>1, i\neq k$); at this level of approximation $\Phi_\mathrm{d}=0$ while $\Phi_\mathrm{id}=-Gm_k/R_k$, and the classical expression for the angular frequency of the two-body problem is easily recovered. Once this is done, at the next level of approximation, one would include all other terms proportional to $m_i=0$ ($i>1, i\neq k$).


\section{Conservation laws in a non-inertial frame}
\label{sec:cons}


The advantage of the indirect force interpretation suggested in the previous section becomes particularly clear when considering conservation of momentum, angular momentum and energy in inertial and non-inertial frames. 

In an inertial frame the total momentum ${\bf p}$, angular momentum ${\bf l}$ and energy $e$ of the system of all $N$ bodies defined as 
\begin{align}
{\bf p} & = \sum\limits_{n=1}^N m_n \dot{\bf r}_n\,,
~~~~~~{\bf l} = \sum\limits_{n=1}^N m_n {\bf r}_n\times \dot{\bf r}_n\,,
\label{eq:PLin}
\\
e & = \frac{1}{2}\sum\limits_{n=1}^N m_n \dot{\bf r}_n^2 -\frac{1}{2}\sum\limits_{n=1}^N\sum\limits_{\substack{k=1 \\ k\neq n}}^N \frac{G m_n m_k}{\vert {\bf r}_n-{\bf r}_k\vert}\, ,
\label{eq:Ein}
\end{align}
in terms of coordinates ${\bf r}_n$ in the inertial frame are conserved, i.e.
\ba
\dot {\bf p}={\bf 0},~~~\dot {\bf l}={\bf 0},~~~\dot e=0.
\label{eq:cons_in}
\ea

In a non-inertial system attached to $m_1$, it is natural to define the total momentum ${\bf P}$, angular momentum ${\bf L}$ and energy $E$ as close as possible to the definitions (\ref{eq:PLin})-(\ref{eq:Ein}) but through the coordinates $\bfR_n$ in that frame:
\begin{align}
{\bf P} & = \sum\limits_{n=1}^N m_n \dot{\bf R}_n\,,
~~~~~~ {\bf L} = \sum\limits_{n=1}^N m_n {\bf R}_n\times \dot{\bf R}_n\,,
\label{eq:PLnin}
\\
E & = \frac{1}{2}\sum\limits_{n=1}^N m_n \dot{\bf R}_n^2 -\frac{1}{2}\sum\limits_{n=1}^N\sum\limits_{\substack{k=1 \\ k\neq n}}^N \frac{G m_n m_k}{\vert {\bf R}_n-{\bf R}_k\vert}\,.
\label{eq:Enin}
\end{align}
This makes our definitions of these physical quantities independent of the nature of the coordinate system, be it inertial or non-inertial. Note that since $\bfR_1={\bf 0}$, $\dot\bfR_1={\bf 0}$ by definition, we can also re-write equations (\ref{eq:PLnin})-(\ref{eq:Enin}) as
\begin{align}
{\bf P} & = \sum\limits_{n=2}^N m_n \dot{\bf R}_n\,,
~~~~~~ {\bf L} = \sum\limits_{n=2}^N m_n {\bf R}_n\times \dot{\bf R}_n\,,
\label{eq:PLnin1}
\\
E &= \frac{1}{2}\sum\limits_{n=2}^N m_n \dot\bfR_n^2 -\sum\limits_{n=2}^N \frac{G m_1 m_n}{R_n}-\frac{1}{2}\sum\limits_{n=2}^N\sum\limits_{\substack{k=2 \\ k\neq n}}^N \frac{G m_n m_k}{\vert \bfR_n-\bfR_k\vert}\,,
\label{eq:Enin1}
\end{align}
where we also explicitly separated the gravitational effect of $m_1$ in the definition of $E$.

Let us next define $M=\sum_{k=1}^N m_n$ to be the total mass of all objects in the system and 
\ba
{\bf R}_\mathrm{b}=\frac{1}{M}\sum\limits_{n=1}^N  m_n {\bf R}_n=\frac{1}{M}\sum\limits_{n=2}^N  m_n {\bf R}_n
\label{eq:Rb}
\ea  
to be the barycenter of the system of all $N$ masses in the non-inertial frame (with $\bfR_1={\bf 0}$). With the definition (\ref{eq:R}), one can easily see that ${\bf R}_\mathrm{b}={\bf r}_\mathrm{b}-{\bf r}_1$, where ${\bf r}_\mathrm{b}=M^{-1}\sum_{k=1}^N m_n{\bf r}_n$ is the barycenter of the whole system in the inertial frame.

In Appendix \ref{sec:cons-laws} we derive the following equations for the evolution of  ${\bf P}$, ${\bf L}$, and $E$ in a non-inertial frame:
\ba
&& \dot {\bf P}  = M \aid\, ,
\label{eq:Pcons}\\
&& \dot {\bf L}  = \bfR_\mathrm{b}\times \left(M \aid\right),
\label{eq:Lcons}\\
&& \dot E =
\dot\bfR_\mathrm{b}\cdot \left(M\aid\right).
\label{eq:Econs}
\ea
If the reference frame is in fact inertial and $\aid={\bf 0}$, these equations naturally reduce to equation (\ref{eq:cons_in}).

The equations (\ref{eq:Pcons})-(\ref{eq:Econs}) imply that ${\bf P}$, ${\bf L}$, and $E$ are not conserved in a non-inertial frame. But their form makes it obvious why this is the case. The total momentum ${\bf P}$ changes because in a non-inertial, accelerated frame fixed on $m_1$ the whole system is subject to a reflex acceleration $\aid$, which acts uniformly on all objects and gives rise to an external force $M \aid$ acting on the whole system. According to Newton's second law, this results in the change of the total momentum of the system given by equation (\ref{eq:Pcons}).

Similarly, total angular momentum ${\bf L}$ is not conserved because (uniform) reflex acceleration $\aid$ results in a torque  $\bfR_n\times(m_n\aid)$ acting on each mass $m_n$ in the system. Summed over all objects and with the definition (\ref{eq:Rb}), this torque is found to be equal to the right-hand side of equation (\ref{eq:Lcons}).  

Finally, the energy $E$ changes because the force $m_n\aid$ associated with the reflex acceleration $\aid$ of every mass $m_n$ does work $m_n\dot\bfR_n\cdot\aid$ on this mass, resulting in a change of its energy. When summed up over all objects in the system, this work ends up being equal to the right-hand side of equation (\ref{eq:Econs}).


\section{Two-body problem}
\label{sec:2body}


The two-body problem is a particular case of a general $N$-body problem when $N=2$. It is integrable, and the easiest way to find a solution is via considering the motion of one of the objects relative to another, e.g. $m_2$ relative to $m_1$ with the separation $\bfR_2={\bf r}_2-{\bf r}_1$. Then the relative motion is described by equation (\ref{eq:EoM2body}), which can be integrated once to deduce the existence of the {\it vis viva} integral of motion (IoM)
\begin{align}
I=\frac{1}{2}\dot\bfR_2^2-\frac{G(m_1+m_2)}{R_2}=\frac{1}{2}\dot\bfR_2^2+\Phi_\mathrm{2b}(\bfR_2),~~~~\dot I=0.
\label{eq:IE}
\end{align}
The vis viva integral is commonly interpreted as the `energy integral' of motion in the potential $\Phi_\mathrm{2b}$, implying that the orbital energy per unit mass is conserved \citep{Murray1999,Kop2011}. Indeed, one can show that $\mu I$, where $\mu=m_1m_2/(m_1+m_2)$ is the reduced mass, represents the energy of a two-body system computed in the inertial, barycentric frame \citep{Murray1999,Tremaine2023}. 

However, the vis viva integral is often derived by considering relative motion of two objects in a non-inertial frame attached to one of them \citep{Murray1999}. In that frame the interpretation of $I$ as 
the 'energy integral' is rather misleading. First, as we argued in Section \ref{sec:indirect}, $\Phi_\mathrm{2b}$ does not represent a true gravitational potential of the system and raises issues with the equivalence principle. It should not be considered as the gravitational potential energy of the system.

Second, the extension of the definition of energy $E$ to the case of $N>2$ based on the $N=2$ expression for $I$ (i.e. interpreting $m_n\Phi_\mathrm{2b}$ as a potential energy of gravitational coupling with the mass $m_1$ instead of the true potential energy $m_n\Phi_\mathrm{N}$) is not obvious. Indeed, it does not seem possible to formulate a physically-motivated analog of equation (\ref{eq:Enin}) that would involve $\Phi_\mathrm{2b}$. 

Third, equation (\ref{eq:IE}) simply does not represent the total energy of the relative motion of the two-body system, i.e. the motion of $m_2$ as viewed from a frame co-moving with $m_1$. The true energy $E$ --- the sum of kinetic and gravitational potential energies in the frame of $m_1$--- is given by the equation (\ref{eq:Enin1}) for $N=2$, namely 
\ba
E = \frac{1}{2}m_2 \dot\bfR_2^2 + m_2\Phi_\mathrm{N} = \frac{1}{2}m_2 \dot\bfR_2^2 - \frac{G m_1 m_2}{R_2}\,.
\label{eq:Enin2}
\ea
It is clear that $E$ is different from $I$, even when the latter is multiplied by $m_2$. But this is hardly surprising if one resorts to our interpretation of the relative motion in a non-inertial frame advanced in Section \ref{sec:alt-motion}. 

Indeed, for $N=2$ the indirect acceleration (\ref{eq:EoMRgen-alt1}) becomes 
\ba
\aid = - Gm_2\frac{\bfR_2}{R_2^3}.
\label{eq:acc2}
\ea  
Also, for the two-body system $M=m_1+m_2$ and $\bfR_\mathrm{b}=(m_2/M)\bfR_2$. Plugging this into the right-hand side of equation (\ref{eq:Econs}), one finds that $\dot E$ is non-zero,
\ba
\dot E = - Gm_2^2\frac{\bfR_2\cdot \dot\bfR_2}{R_2^3},
\label{eq:Edot2}
\ea 
so that the energy of the relative motion $E$ is {\it not} an integral of motion. This equation reflects the {\it non-conservation} of $E$ due to the work done on the two-body system by the indirect force. Note that in our interpretation $E$ is defined with the true gravitational potential $\Phi_\mathrm{N}$ satisfying the equivalence principle, but unlike the classical two-body interpretation (see Section \ref{sec:indirect}), there is an additional force in the system --- the indirect force.

Given its simple form for $N=2$, equation (\ref{eq:Edot2}) can be easily integrated in time, resulting in an integral of motion 
\ba
E - \frac{Gm_2^2}{R_2}=\frac{1}{2}m_2 \dot\bfR_n^2 - \frac{G m_1 m_2}{R_2}- \frac{Gm_2^2}{R_2}=m_2I,
\label{eq:IE1}
\ea  
where we used equations (\ref{eq:IE}) and (\ref{eq:Enin2}). In other words, in a non-inertial frame the IoM $m_2I$ is only the `energy-like' IoM that represents the conservation of the {\it sum} of the total energy of the system $E$ and the work done upon it by the indirect force $M\aid$. This analysis illustrates why calling $I$ (or $m_2I$) the `energy integral' can be misleading when considering motion in the frame of one of the bodies. Using this specific name in relation to $I$ in that frame should be avoided.

Turning now to the angular momentum of the system in a non-inertial frame, we confirm a well-known fact that ${\bf L}$ is conserved and represents another IoM. This follows from equation (\ref{eq:Lcons}) and the fact that $\bfR_\mathrm{b}$ and $\aid$ are aligned in a two-body case.

Finally, equation (\ref{eq:Pcons}) for momentum ${\bf P}=m_2\dot\bfR_2$ evolution
\ba
\frac{\de}{\de t}\left(m_2\dot\bfR_2\right)=- Gm_2\frac{\bfR_2}{R_2^3}\times (m_1+m_2)
\ea
is mathematically identical to equation (\ref{eq:EoM2body}) multiplied by $m_2$. However, the interpretation of the right-hand side --- the force --- is different in two cases: in equation (\ref{eq:EoM2body}) it represents the action of the fictitious two-body potential $\Phi_\mathrm{2b}$, whereas in equation (\ref{eq:Pcons}) the right-hand side is the full indirect force acting on the system, which arises naturally when the frame of $m_1$ is accelerated by $m_2$.


\section{Discussion}
\label{sec:disc}


One of the main goals of this work was to highlight the benefits of using the modified form (\ref{eq:acc}) of the indirect acceleration $\aid$ in comparison with the standard form (\ref{eq:indirect-pot}) conventionally used in celestial mechanics. It should be emphasized that using this form of $\aid$ does not in any way change the actual calculations of motion of objects in $N$-body systems. The laws of motion given by equations (\ref{eq:EoMRgen}) and (\ref{eq:EoMRgen-alt1}) are identical to each other, regardless of how we partition their different terms.

Nevertheless, introduction of $\aid$ via the equation (\ref{eq:acc}) allows us to provide a simple and clear {\it interpretation} of the dynamics in a non-inertial system. In particular, we showed that by using $\aid$ one can formulate a simple and intuitive set of conservation laws (see equations (\ref{eq:Pcons})-(\ref{eq:Econs})) for the total momentum, angular momentum and energy of the system defined in a pretty standard and physically-motivated form (see equations (\ref{eq:PLnin1})-(\ref{eq:Enin1})) even in a non-inertial frame. These conservation laws state that ${\bf P}$, ${\bf L}$, and $E$ do not remain constant but change as a result of indirect force $M\aid$ acting on the full system and its associated torque and work. In turn, this force arises because the indirect acceleration $\aid$ acts on all bodies in the system in the same way and is considered as an external agent acting on the system. As a result, the conservation laws (\ref{eq:Pcons})-(\ref{eq:Econs}) have the same meaning as they would have in an inertial frame, in the presence of an external force acting on the system of $N$ bodies.

This interpretation is to be contrasted with e.g. the classical `conservation law' (\ref{eq:IE}) for a two-body system in a non-inertial frame, that requires (i) the introduction of a special 2-body potential $\Phi_\mathrm{2b}$ that does not obey the equivalence principle and (ii) the definition of `energy' $I$ that cannot be easily extended to systems of $N>2$ objects.  

Equations (\ref{eq:Pcons})-(\ref{eq:Econs}) illustrate the instantaneous non-conservation of ${\bf P}$, ${\bf L}$, and $E$ in a non-inertial frame. However, over timescales long compared to the dynamical timescale of the system, their right-hand sides should average out to zero, leading to the usual conservation of the total energy and angular momentum of systems of planets in a secular approximation (at least as long as these systems remain finite and bound). In fact, indirect forces are known to have no effect in secular perturbation theory in celestial mechanics \citep{Murray1999}.

Interestingly, standard tests of the performance of various integration algorithms in celestial mechanics are often focused on integrating the two-body problem and observing the conservation of energy in an inertial, barycentric frame (which is the same as the vis-viva integral $I$, up to a constant factor). Since this frame is different from a non-inertial frame attached to one of the objects, the short-term energy non-conservation described by equation (\ref{eq:Econs}) is not observed in such tests.

Understanding indirect forces is important not only in celestial mechanics. For example, they also arise in studies of disc-planet interaction \citet{GT80}, for which a natural coordinate system is the one attached to the central star. \citet{Crida2025} and \citet{Rafikov2025} explored the role played by the indirect force in disc-planet studies, finding it to be important in situations where the planet is massive enough to significantly perturb the disc and open a gap around its orbit. \citet{Rafikov2025} also examined global angular momentum balance of the disc-planet system in the non-inertial stellar frame, showing that the full angular momentum of the system is not conserved. This finding is fully in line with our equation (\ref{eq:Lcons}) demonstrating non-conservation of ${\bf L}$ in non-inertial frames.


\section{Summary}
\label{sec:sum}


In this work we provided a non-conventional (from the celestial mechanics perspective) interpretation of the dynamics in non-inertial reference frames, applicable to $N$-body systems such as planetary systems. By defining the indirect acceleration in a way slightly different from the conventional practice, one is able to cast the conservation laws of momentum ${\bf P}$, angular momentum ${\bf L}$ and energy $E$ (defined in a physically-motivated fashion) in a non-inertial frame in a straightforward and intuitive form. 

In this approach the indirect acceleration ends up being the same for all objects of the system (which is different from the standard convention adopted in celestial mechanics) and affects the system's dynamics as an external force, leading to instantaneous non-conservation of ${\bf P}$, ${\bf L}$, and $E$. This non-conservation has important implications not only for $N$-body systems but also for disc-planet interaction --- gravitational coupling between a massive planet and a fluid disc, as we show in \citet{Rafikov2025}. 

We also demonstrate that the vis-viva integral of the two-body problem, which is classically considered to reflect energy conservation for the relative motion of two gravitating masses, in our interpretation describes energy non-conservation in a non-inertial frame: the total energy associated with the relative motion of two gravitating objects in that frame changes due to the work done by the indirect force. Thus, the vis-viva integral is only an energy-like integral of motion in a non-inertial frame attached to one of the objects, and not the true energy (sum of kinetic and potential energies) of relative motion in a two-body system.


\section*{Acknowledgements}


The author is grateful to Callum Fairbairn and Alexander Dittmann for their comments on the manuscript, to Scott Tremaine and Nicolas Cimerman for useful discussions, and to Daniel Tamayo and an anonymous referee for their suggestions that helped sharpen the logic of the manuscript. R.R.R. acknowledges financial support from the STFC grant ST/T00049X/1  and the IBM Einstein Fellowship at the IAS.


\section*{Data availability}
No new data were generated or analysed in support of this research.



\bibliographystyle{mnras}
\bibliography{references} 



\appendix


\section{Conservation laws in the non-inertial frame}
\label{sec:cons-laws}


Here we derive the conservation laws given by equations (\ref{eq:Pcons})-(\ref{eq:Econs}) in a non-inertial frame. These derivations hold regardless of the interpretation of indirect acceleration that is used (see Sections \ref{sec:indirect} \& \ref{sec:alt-motion}), as the equations of motion are the same in both cases, only the partition of different terms varies. Despite that,  in what follows we find it more convenient to operate with equation (\ref{eq:EoMRgen-alt1}) rather than equation (\ref{eq:EoMRgen}).


\subsection{Momentum conservation}
\label{sec:cons-laws-p}


Differentiating the expression (\ref{eq:PLnin}) for ${\bf P}$ with respect to time and using equation (\ref{eq:EoMRgen-alt1}), we find
\begin{align}
\dot {\bf P}  = \sum\limits_{n=2}^N m_n \ddot{\bf R}_n &= 
 -m_1\sum\limits_{n=2}^N Gm_n\frac{{\bf R}_n}{R_n^3}
 \nonumber\\
 & -
G\sum\limits_{n=2}^N \sum\limits_{\substack{k=2 \\ k\neq n}}^N m_n m_k\frac{{\bf R}_n-{\bf R}_k}{\vert{\bf R}_n-{\bf R}_k\vert^3}+\aid\sum\limits_{n=2}^N m_n.
\label{eq:Pderiv}
\end{align}
One can see that the second term in the right-hand side is zero because of anti-symmetry of each individual term in the double sum with respect to indices $k$ and $n$ and our freedom to change the order of summation in the double sum. Also, in light of the definition (\ref{eq:acc}), the first term in the right-hand side is simply $m_1\aid$; combining it with the last term in the right-hand side and using the definition of $M$, one arrives at equation (\ref{eq:Pcons}).


\subsection{Angular momentum conservation}
\label{sec:cons-laws-l}


Differentiating now the expression (\ref{eq:PLnin}) for ${\bf L}$ with respect to $t$ and again using equation (\ref{eq:EoMRgen-alt1}), we find
\begin{align}
& \dot {\bf L}=\sum\limits_{n=2}^N m_n \frac{\de}{\de t}\left({\bf R}_n\times \dot{\bf R}_n\right) =\sum\limits_{n=2}^N m_n ~{\bf R}_n\times \ddot{\bf R}_n 
\nonumber\\
& =G\sum\limits_{n=2}^N \sum\limits_{\substack{k=2 \\ k\neq n}}^N m_n m_k
\frac{{\bf R}_n\times{\bf R}_k}{\vert{\bf R}_n-{\bf R}_k\vert^3}+\sum\limits_{n=2}^N m_n {\bf R}_n\times \aid \,,
\label{eq:fullL1}
\end{align}
Once again, the first term in the right-hand side (direct contribution) goes away due to antisymmetry across the $n$, $k$ indices and freedom to change the order of summation. The remaining second (indirect) part can be written as (remembering that $\bfR_1={\bf 0}$)
\begin{align}
\dot {\bf L}=\left(\frac{1}{M}\sum\limits_{n=1}^N m_n {\bf R}_n\right)\times \left(M \aid\right)
\label{eq:fullL2}
\end{align}
which, after using definition (\ref{eq:Rb}), reduces to the equation (\ref{eq:Lcons}).


\subsection{Energy conservation}
\label{sec:cons-laws-e}


To derive equation (\ref{eq:Econs}) we first take the time derivative of the definition (\ref{eq:Enin1}):
\begin{align}
\dot E &= \sum\limits_{n=2}^N m_n ~\dot\bfR_n\cdot \ddot\bfR_n +\sum\limits_{n=2}^N G m_1 m_n\frac{\bfR_n\cdot\dot\bfR_n}{R_n^3}
\nonumber\\
& +\frac{1}{2}\sum\limits_{n=2}^N\sum\limits_{\substack{k=2 \\ k\neq n}}^N G m_n m_k\frac{(\bfR_n-\bfR_k)\cdot(\dot\bfR_n-\dot\bfR_k)}{\vert \bfR_n-\bfR_k\vert^3}\,,
\label{eq:Eeq1}
\end{align}
Upon substituting $\ddot\bfR_n$ from equation (\ref{eq:EoMRgen-alt1}), the first term in the right-hand side of (\ref{eq:Eeq1}) becomes
\begin{align}
&-\sum\limits_{n=2}^N G m_1 m_n\frac{\bfR_n\cdot\dot\bfR_n}{R_n^3}-\sum\limits_{n=2}^N\sum\limits_{\substack{k=2 \\ k\neq n}}^N G m_n m_k\frac{\dot\bfR_n\cdot(\bfR_n-\bfR_k)}{\vert \bfR_n-\bfR_k\vert^3}\nonumber\\
& +\sum\limits_{n=2}^N m_n ~\dot\bfR_n\cdot \aid\,,
\label{eq:Eeq2}
\end{align}
Plugging this back into (\ref{eq:Eeq1}), after obvious cancellations of different terms, one finds 
\begin{align}
\dot E &=-\frac{1}{2}\sum\limits_{n=2}^N\sum\limits_{\substack{k=2 \\ k\neq n}}^N G m_n m_k\frac{(\bfR_n-\bfR_k)\cdot(\dot\bfR_n+\dot\bfR_k)}{\vert \bfR_n-\bfR_k\vert^3}
\nonumber\\
&+\sum\limits_{n=2}^N m_n ~\dot\bfR_n\cdot \aid\,,
\label{eq:Eeq3}
\end{align}
The double sum on the right-hand side again vanishes because of anti-symmetry with respect to $k$, $n$ indices (and symmetry in changing the order of summation), so, remembering that $\dot\bfR_1={\bf 0}$, we are left with
\begin{align}
\dot E = \left(\frac{1}{M}\sum\limits_{n=1}^N m_n ~\dot\bfR_n\right)\cdot \left(M\aid\right)\,.
\end{align}
In light of definition (\ref{eq:Rb}), this expression reduces to equation (\ref{eq:Econs}).


\bsp	
\label{lastpage}
\end{document}